\documentclass[english,twocolumn,prl]{revtex4}
\usepackage[T1]{fontenc}
\usepackage{bm}
\usepackage{amsmath}
\usepackage{amssymb}
\usepackage{lipsum}
\usepackage{graphicx}
\usepackage{bm}
\usepackage{ dsfont }
\usepackage{makeidx}
\usepackage{epstopdf} 
\usepackage[retainorgcmds]{IEEEtrantools}
\usepackage{hyperref} 
\usepackage{braket}

\newcommand{\beq}{\begin{equation}} 
\newcommand{\eeq}{\end{equation}} 
\newcommand{\beqa}{\begin{eqnarray}} 
\newcommand{\eeqa}{\end{eqnarray}}

\makeatletter

\@ifundefined{textcolor}{}
{%
 \definecolor{BLACK}{gray}{0}
 \definecolor{WHITE}{gray}{1}
 \definecolor{RED}{rgb}{1,0,0}
 \definecolor{GREEN}{rgb}{0,1,0}
 \definecolor{BLUE}{rgb}{0,0,1}
 \definecolor{CYAN}{cmyk}{1,0,0,0}
 \definecolor{MAGENTA}{cmyk}{0,1,0,0}
 \definecolor{YELLOW}{cmyk}{0,0,1,0}
 }

\makeatother

\begin{document}
\bibliographystyle{naturemag}
\title{ Robust band of critical states in T broken fermionic systems with lattice selective
disorder}

\author{ Eduardo V. Castro}
\email{eduardo.castro@tecnico.ulisboa.pt}
\affiliation{Centro de F\'isica das Universidades do Minho e Porto,
Departamento de F\'isica e Astronomia, Faculdade de Ciencias,
Universidade do Porto, 4169-007 Porto, Portugal}
\affiliation{CeFEMA, Instituto Superior T\'{e}cnico, Universidade de 
Lisboa, Av. Rovisco Pais, 1049-001 Lisboa, Portugal}
\affiliation{Beijing Computational Science Research Center, Beijing 100084, China }

\author{ Raphael de Gail}
\affiliation{Instituto de Ciencia de Materiales de Madrid, CSIC,
Sor Juana In\'es de la Cruz 3,  Cantoblanco,
E-28049 Madrid, Spain}
\author{M. Pilar L\'opez-Sancho}
\affiliation{Instituto de Ciencia de Materiales de Madrid, CSIC,
Sor Juana In\'es de la Cruz 3,  Cantoblanco,
E-28049 Madrid, Spain}

\author{Mar\'{\i}a A. H. Vozmediano}
\affiliation{Instituto de Ciencia de Materiales de Madrid, CSIC,
Sor Juana In\'es de la Cruz 3,  Cantoblanco,
E-28049 Madrid, Spain}

\begin{abstract}
We analyze the localization properties of  two dimensional systems based on partite lattices with a basis.  
Contrary to standard results, we find that a band of critical states emerges for systems in the unitary class A 
preserving spin symmetry when disorder is unevenly distributed over the basis atoms. The critical metal arises when the less disordered  sublattice
is connected and has time reversal symmetry broken. The
unexpected robustness to disorder presented here is an appealing result which can be measured in
optical lattices.
\end{abstract}

\maketitle

\section{ Introduction} 
\label{sec_intro}
Anderson localization \cite{Anderson58} in one of the best studied phenomena in modern condensed matter. 
The localization transition in d = 2 depends
critically on the symmetry class of the system \cite{AALR79}.  In the standard symmetry classification completed with the addition of topological properties \cite{SRFL08}, the behavior of the unitary class (class A)  was initially understood in the context of the quantum Hall effect where the spin degree of freedom is still
a good quantum number.  It was established
numerically that at the center of each Landau
level band there is only one critical state -- an extended
state where the localization length diverges linearly with
system size --, indicative of a vanishing $\beta$ scaling function \cite{MK81,MK83}.
The case when both time reversal symmetry ${\cal T}$ and spin
rotation symmetry are broken has been the subject of
recent investigation \cite{QHetal16,SWetal16,WSetal15,XSetal12}, and the physics was found
to be different. It seems well established that a band of
extended states, and not a single state, shows up. Depending
on the model, this band can be made entirely
of critical states \cite{SWetal16,XSetal12}, or can be a band of truly extended
states \cite{QHetal16,SWetal16}. Interestingly enough, the transition
from the localized side to the critical metal at the energy
$E_c$ is accompanied by a divergent localization length
$\xi(E)$ of the form 
$\xi(E)\sim \exp (\alpha/\sqrt{\mid E-E_c\mid})$, reminiscent
of a Berezinskii-Kosterlitz-Thouless transition \cite{K74}.
Critical metallic behavior has also been found in Weyl semimetals \cite{SPBB14,CSetal15,CCetal18}.

In this work we show examples of critical metallic behavior in class A systems with spin rotation symmetry. The critical metal is found on lattices with a basis when disorder is unevenly distributed over the basis atoms. The minimal conditions over the less disordered sublattice for the occurrence of the critical metal are:  It has to be connected and have broken time reversal symmetry.

In all cases we will have a Hamiltonian defined on a lattice with two or more sublattices. Potential (Anderson) disorder is implemented by adding to the Hamiltonian the term $\sum_{i\in A,B}\varepsilon_{i}c_{i}^{\dagger}c_{i}$,
with a uniform distribution of random local energies, $\varepsilon_{i}\in[-W/2,W/2]$. 
For selective disorder we have disorder strength $W_i$ for sublattice $i$.

\section{ Models and results }
We will begin by summarizing the situation of the Haldane model \cite{H88}, a prototype of class A system with spin rotation symmetry.  
A robust metallic state has already been found by introducing strong selective disorder 
in one sublattice \cite{CLV15,CGL16}. 
The  tight binding Hamiltonian can be written as
\beqa
H  &=  -t\displaystyle\sum_{\langle i,j\rangle}c_i^{\dagger}c_j
-t_2\sum_{\langle\langle i,j\rangle\rangle}e^{-i\phi_{ij}}c_i^{\dagger}c_j
&+  \Delta\displaystyle\sum_i \eta_i c_i^{\dagger}c_i,
\label{TBHmodel}
\eeqa
where $c_i=A,B$ are defined in the two triangular sublattices that form the honeycomb lattice. 
The first term $t$ represents a standard
real nearest neighbor hopping that links the two triangular sublattices.
The $t_2$ term represents a complex next nearest neighbor (NNN) hopping $t_2 e^{-i\phi_{ij}}$
acting within each triangular sublattice with a 
phase  $\phi_{ij}$ that has
opposite signs $\phi_{ij}=\pm\phi$ in the two sublattices.
This term breaks time--reversal symmetry and opens a non--trivial topological gap
at the Dirac points. We have done our calculations for the simplest
case $\phi=\pi/2$. The last term represents a staggered potential ($\eta_i=\pm 1$). It breaks 
inversion symmetry and opens a trivial gap at the Dirac points. We
use spinless fermions; taking spin into account amounts to a spin
degeneracy factor, and lead to the same physics.

We have shown previously \cite{CGL16} that disordering a single
sublattice, i.e. making $W_{A}\neq 0$ but keeping $W_{B} = 0$,
a band of critical states appears for $W_{A} > W^c_{A}$
and it is robust no matter how large is the disorder. This result
has been established using both a level spacing statistics
analysis and the transfer matrix method. 
  
\begin{figure}
\begin{centering}
\includegraphics[width=0.9\columnwidth]{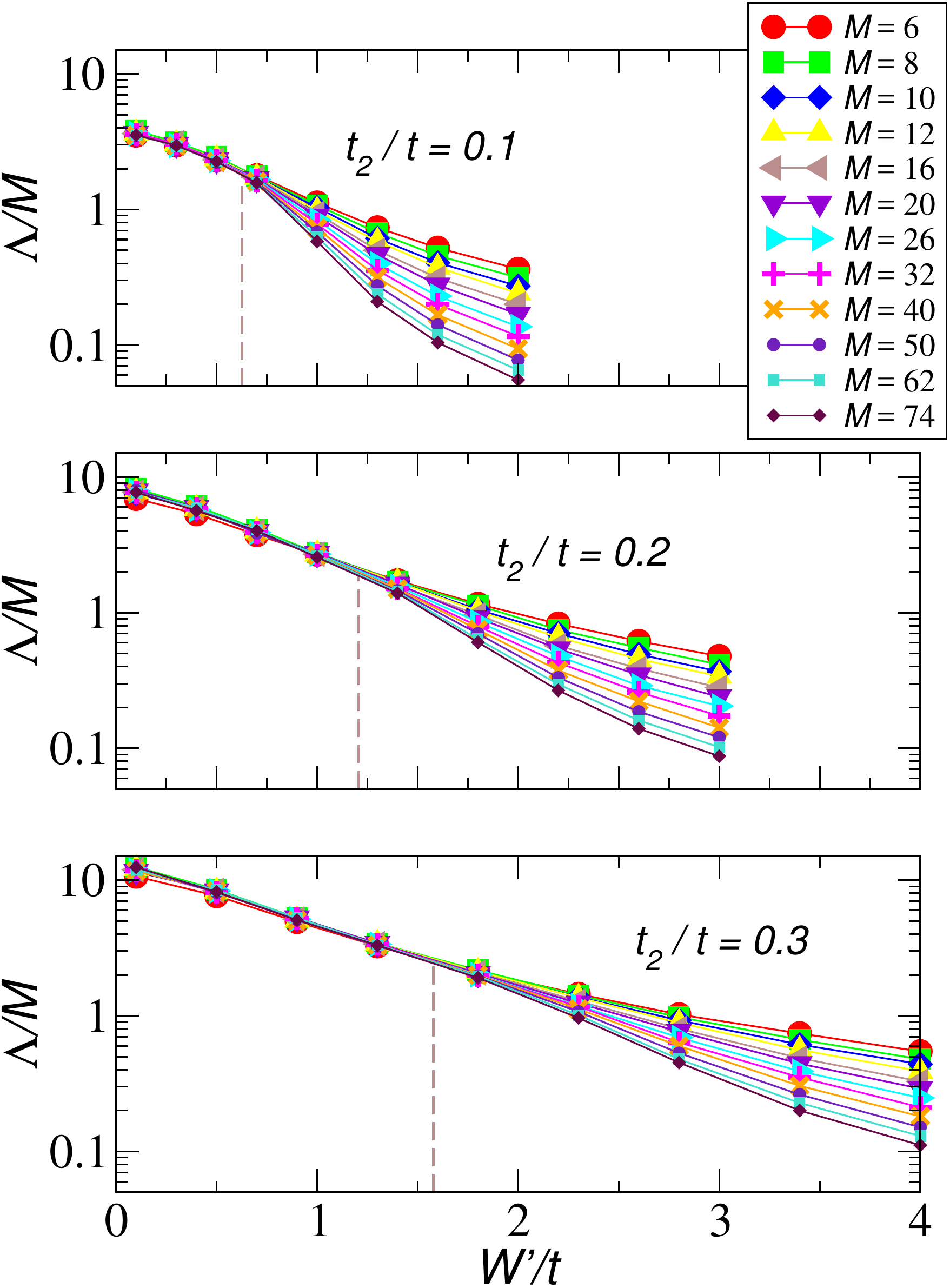}
\par
\end{centering}
\caption{Normalized localization length $\Lambda/M$  as a function of
disorder strength $W_{B}$ in the less disordered (LD) sublattice for various NNN
hoppings $t_2$, calculated for a long tube with M unit cells in
circumference. The vertical dashed line is a guide to the eye
for the transition from critical to localized behaviour. We
have fixed the disorder $W_A=100t$ and chosen the energy $E=-0.05t$.}
\label{fig_loc1}
\end{figure}

In order to explore the robustness
of the metallic state  we study its localization
behavior  with a transfer matrix analysis. We
fix the disorder $W_{A}=100t$ and study the
metal-insulator transition as a function of $W_{B}$ for a
given energy inside the band of critical states. We chose
$E=-0.05t$, and change the disorder strength $W_{B}$. The
results are shown in Figs. \ref{fig_loc1} where we show the normalized
localization length $\lambda/M$ as a function of  disorder strength in the less disordered sublattice,
$W_{LD}$, for
various values of the NNN hopping $t_2$, calculated for a
long tube with M unit cells in circumference. It is obvious
that there is a critical disorder strength $W^c_{LD}$ below
which $\lambda/M$ does not change with M, signalling the presence
of a critical state. For $W_{LD}>W^c_{LD}$, the normalized
localization length decreases with $M$, as expected for a
localized state.
%

%
The results above point to the direction that 2D class A
systems not always turn into an insulator (trivial or nontrivial)
after ``levitation and pair annihilation" has occurred
between extended states. For systems with more
than one sublattice and selective disorder, i.e. at least
one sublattice with low disorder below some critical value
while the remaining atoms have high disorder, the final 
state might be a metal. In what follows  we try to identify the minimal
ingredients needed for such a state to occur. 

The main characteristics that we consider relevant for the physics presented 
in the case of the Haldane model are: 
 1. The original system has bands with non trivial Berry curvature.
 2. After strong selective disorder a ``clean sublattice" with full connectivity remains.
 In the case of the Haldane model it is a triangular lattice with complex NN hopping parameter. 
 3. The remaining``clean sublattice" has complex hopping parameters that break time reversal symmetry irrespective of the possible global topological triviality of the original lattice. 
 
Next we will present alternative lattice models with a richer sublattice structure  
where we can discriminate the necessity of these ingredients.

{\it 1. Honeycomb lattice with enlarged unit cell models.}

%
In ref. \cite{CGL16} we showed that the critical
metal arises in the Haldane model irrespectively
of whether the clean limit phase is trivial (non zero Chern number) or topologically non trivial.
In any case, the triviality in that case was due to a cancellation of the Chern number upon integration of a non--zero Berry curvature. In order to explore the importance of the Berry curvature on the localization properties
of the system we study two different tight--binding
models realized in the honeycomb lattice with a
tripled unit cell. The basis vectors
in real space read $\mathbf{a}_{1}=\frac{3a}{2}(-\sqrt{3},1)$ and 
$\mathbf{a}_{2}=\frac{3a}{2}(\sqrt{3},1)$,
and the respective unit cell vectors in reciprocal space are
$\mathbf{b}_{1}=\frac{2\pi}{3\sqrt{3}a}(-1,\sqrt{3})$ and
$\mathbf{b}_{2}=\frac{2\pi}{3\sqrt{3}a}(1,\sqrt{3})$.
The tight binding Hamiltonian for the two cases we are
interested in, can be generically written as
\begin{equation}
H=-t\sum_{{{\bf r},{\bm \delta}}} a^\dagger_{{\bf r}} b_{{{\bf r} + {\bm \delta}}}
\;+\;
V\sum_{{\bf r}, {\bm \delta}}a^\dagger_{{\bf r }}a_{{\bf r }}
b^\dagger_{{\bf r } + {\bm \delta}}b_{{\bf r } + {\bm \delta}} 
\;+\;
h.c. \;, \label{ham-fs}
\end{equation}
where $t$ is the nearest neighbor hopping and $V$ the nearest 
neighbor Coulomb repulsion. We use standard notation where $a_{{\bf r }}$
($b_{{\bf r }}$) annihilates an electron at position ${\bf r}$ in sublattice
A (B). 
A mean field analysis of the nearest neighbour Hubbard interaction in this  model was studied 
in refs. \cite{CGetal11,GCetal13} to derive spontaneous breaking of time reversal symmetry. 
Two ${\cal T}$ broken phases TI and TII were found associated to the complex NN hopping distributions shown
schematically in fig. \ref{fig_fluxes}. 
\begin{figure}
\begin{centering}
\includegraphics[width=0.8\columnwidth]{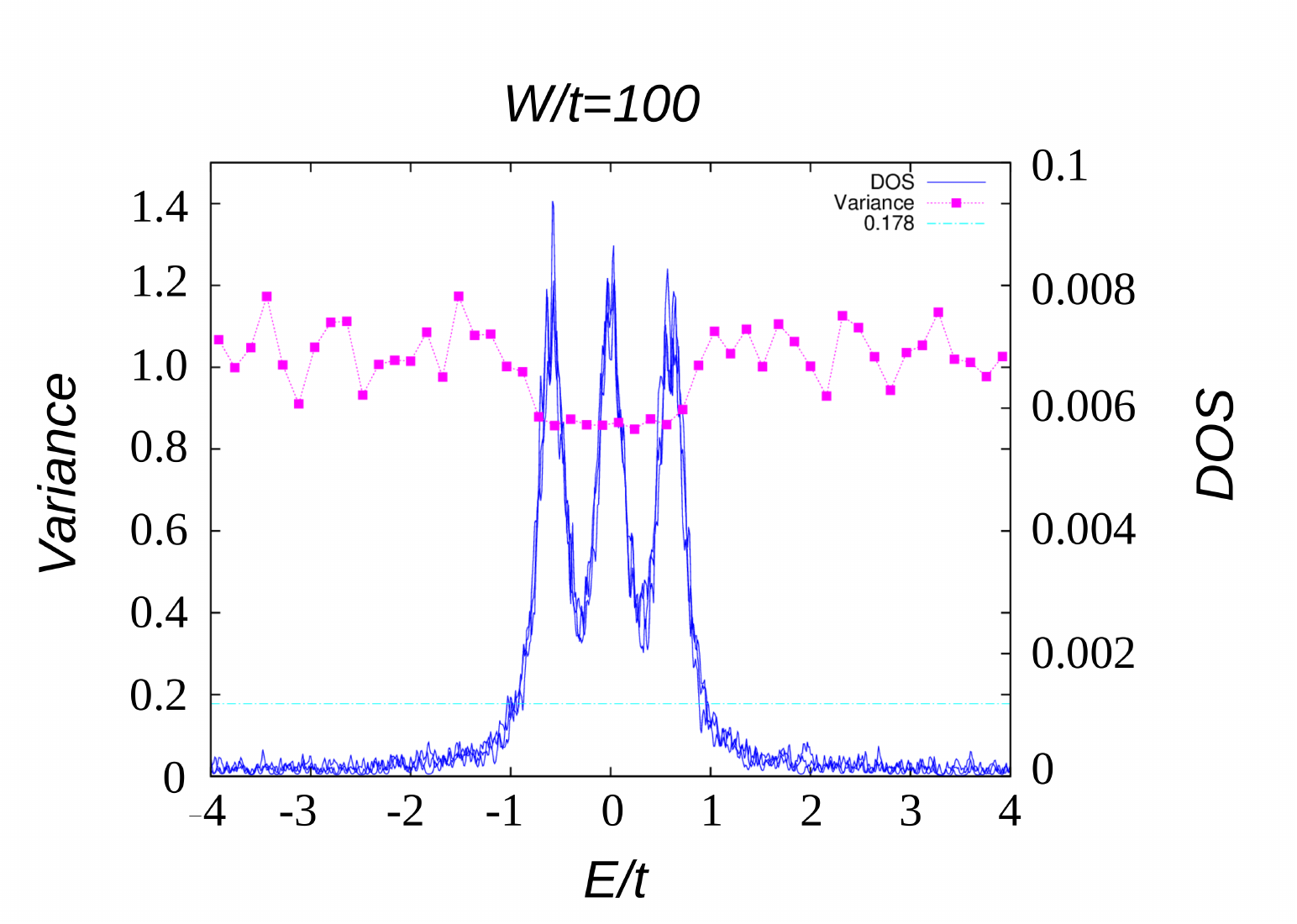}
\par
\end{centering}
\caption{Level spacing variance of the model TI with disorder distributed over three sublattices.
The remaining lattice is not connected and a metallic state does not form.}
\label{fig_varianceTI}
\end{figure}
\begin{figure}
\begin{centering}
\includegraphics[width=0.8\columnwidth]{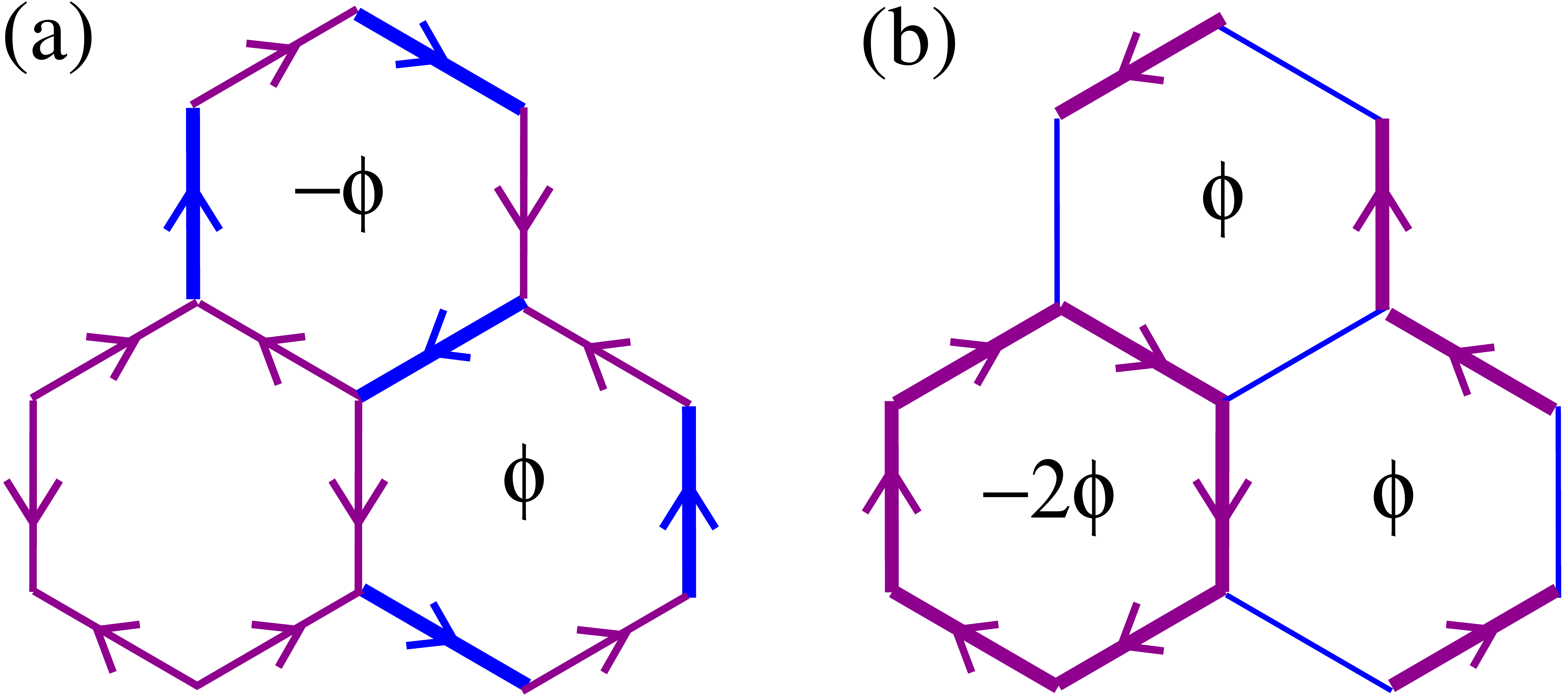}\vspace{0.5cm}
(c)\includegraphics[width=1.0\columnwidth]{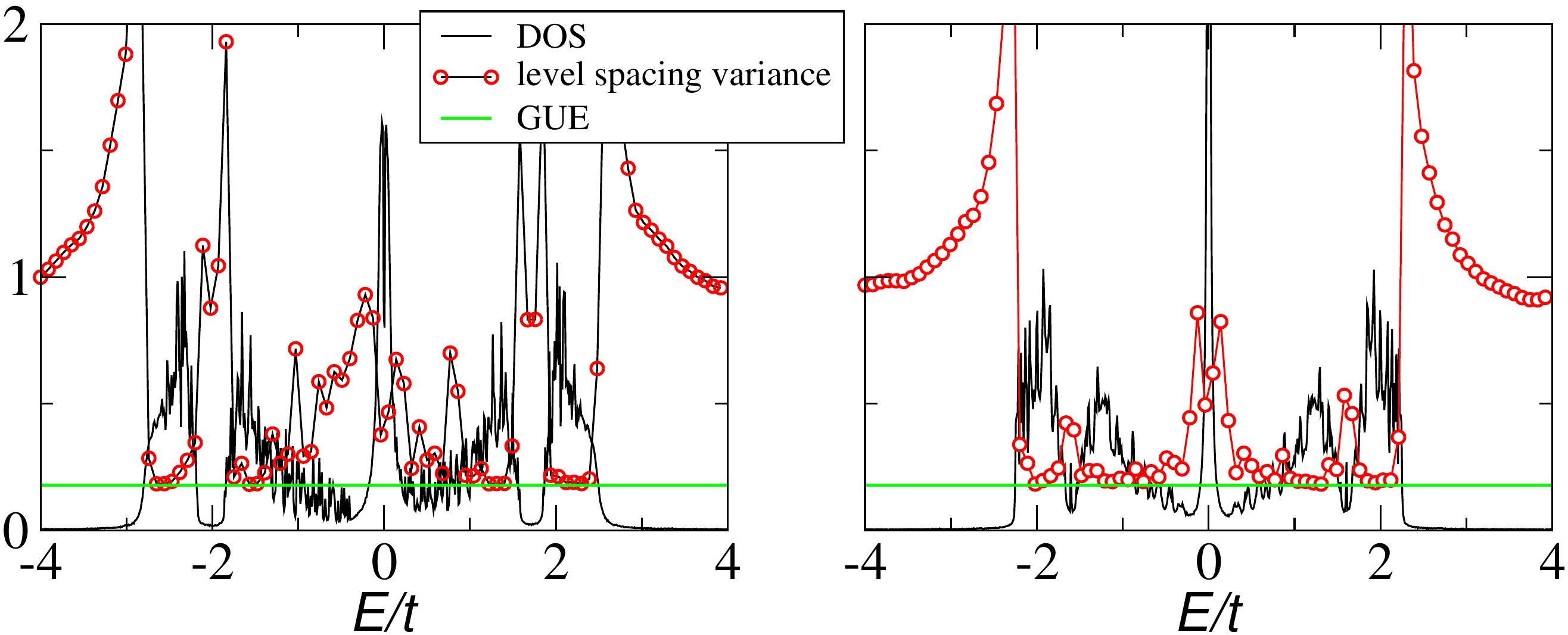}
\par
\end{centering}
\caption{Schematic representation in terms of effective hoppings
and fluxes for phase T-I (a) and TII (b) discussed in the text. (c) Level spacing variance of the model TI  and TII showing the appearance of the critical metal
for strong selective disorder in one of the six sublattices.}
\label{fig_fluxes}
\end{figure}
With respect to the usual tight binding model
for graphene, the T-I phase is just a complex  
renormalization of the bare hopping
plus a Kekule complex distortion. The latter breaks
${\cal T}$ and ${\cal I}$
though preserving their product. According to the analysis in ref. \onlinecite{SF08}
and as shown explicitly in \cite{CGetal11,GCetal13}, the Berry curvature is identically
zero everywhere. 

The bands in the TII phase have non--trivial Berry curvature.
We have used these models first to establish the necessity of the connectivity of the remaining clean lattice
to have an emerging metal. To this purpose, we have studied the level spacing variance 
of two representative Hamiltonians in the phases TI and TII when selective disorder is acting upon three of the six lattice basis, i. e. , we disorder the sublattices $A_i$ equally, leaving
the sublattices $B_i$ clean. This disorder distribution is equivalent to the one performed in the Haldane model. In this case, since there is not NNN hopping parameter, the remaining clean sublattice is not connected. The level space variance of the TI model is shown in Fig. \ref{fig_varianceTI}. As we see, all states are localized. The same occurs with the TII case.

Next, we disorder one of the six sublattices, say
sublattice $A_1$. We use $W = 100t$. The level space variance of both models is shown in the lower part of Fig. \ref{fig_fluxes}.  As we can see, a metallic phase arises for strong selective disorder of one of the six sublattices.
The behavior of the localization length with the size of the lattice points to the fact that,
as happened in the Haldane model, the resulting metal is also a critical metal. We have checked that standard (equally distributed among the two sublattices) Anderson disorder induces normal localization in both lattices. The behavior of the TI lattice shows that a non trivial topology is not a necessary condition to support the metallic phase. The next examples will show that the essential property (shared by both TI and TII lattices)  is the breakdown of time reversal invariance in the ``clean" sublattice.

{\it 2. Square lattice with two orbitals per site and diatomic square lattice.}
\begin{figure}
\begin{centering}
(a)\includegraphics[width=0.40\columnwidth]{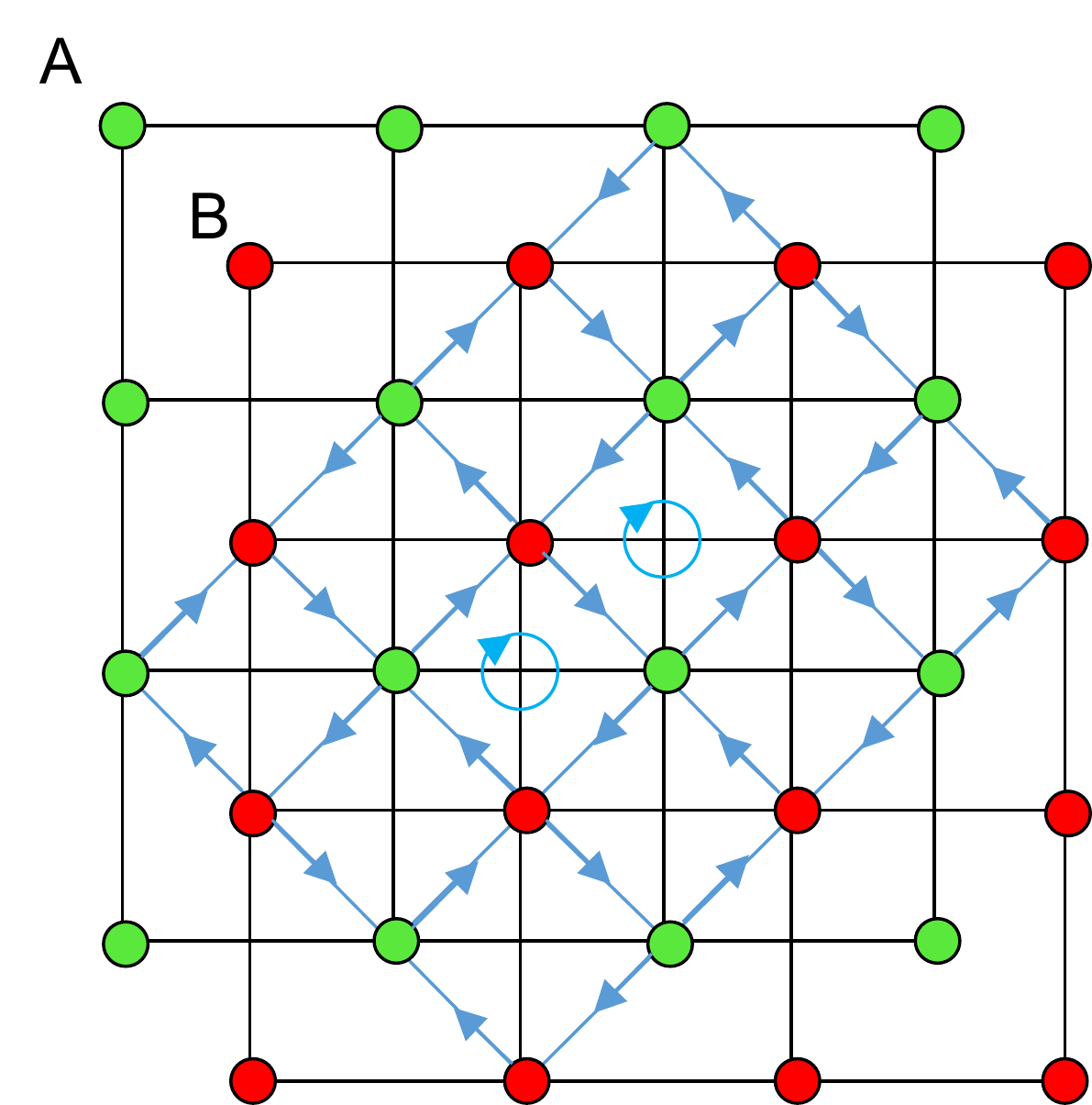} \qquad
(b)\includegraphics[width=0.40\columnwidth]{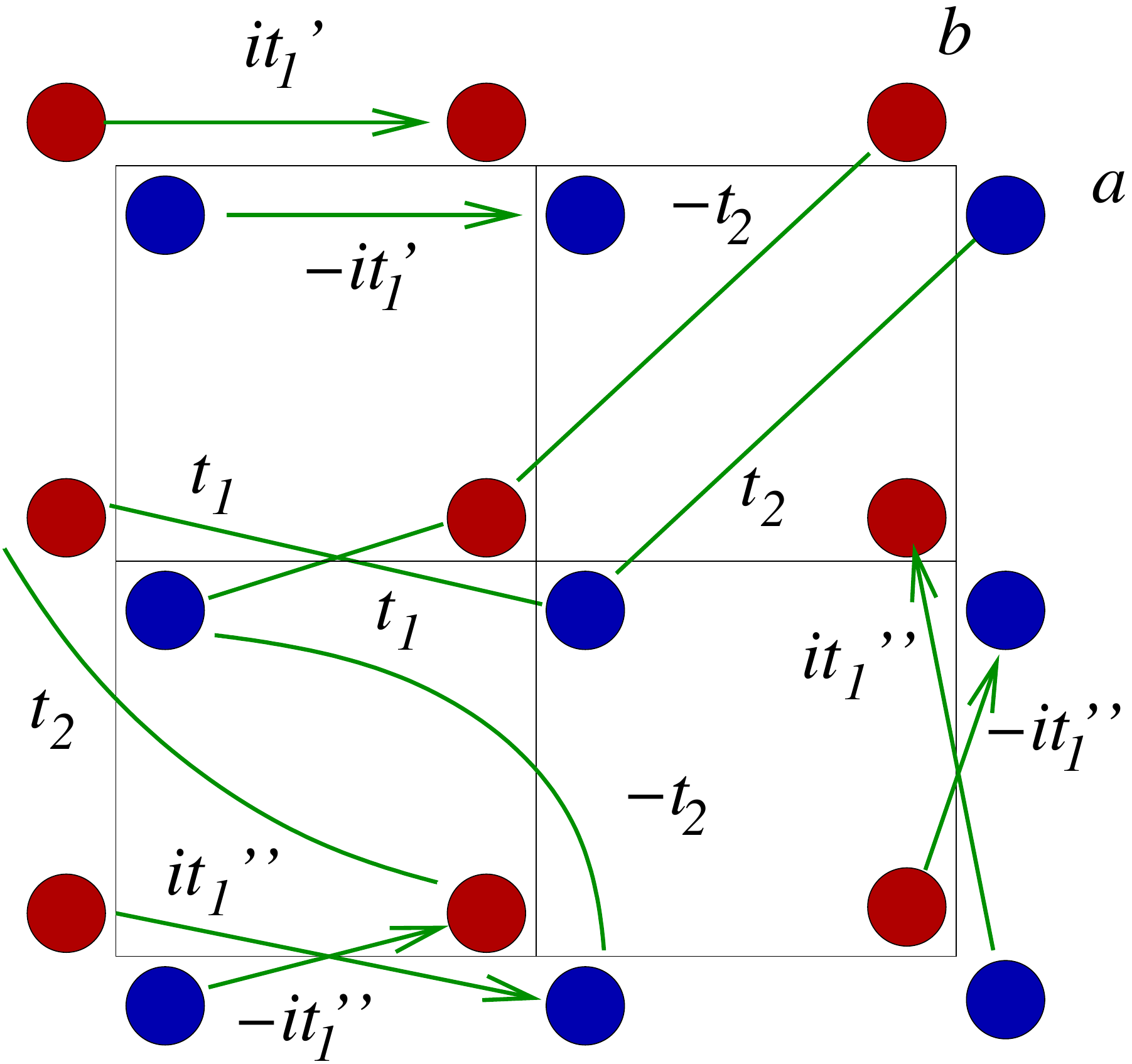}\vspace{0.2cm}
(c)\includegraphics[width=0.90\columnwidth]{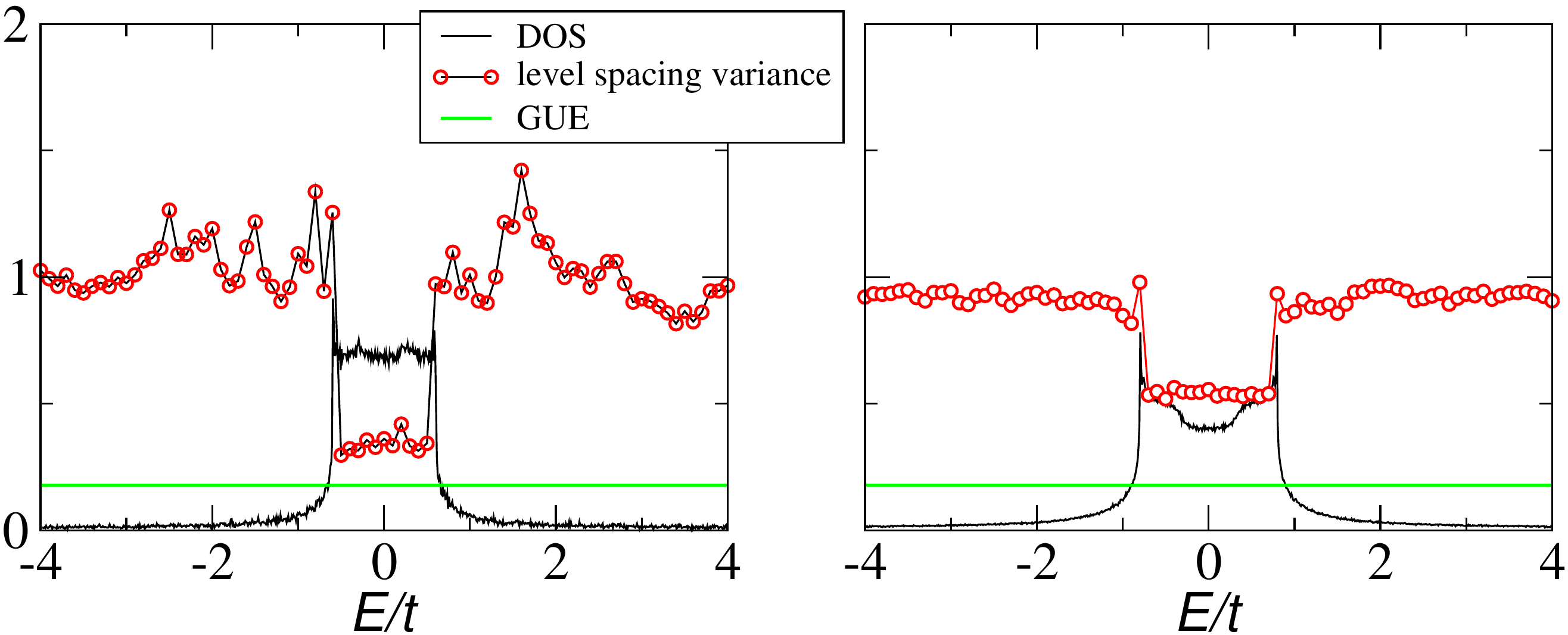}  
\par
\end{centering}
\caption{ Schematic representation of the hopping structure of the diatomic  (a) and two-orbital  model (b) on the square lattice. 
(c) Level spacing variance of the lattices in the upper part showing standard localization for selective disorder in only one sublattice.}
\label{fig_square}
\end{figure} 
Our next examples of a simple Chern insulator are supported on the  square lattice with two internal degrees
of freedom per site that can represent orbital or spin degeneracy $s=\pm$ \cite{WBR12}. Both tight binding models share the property of having complex NN neighbors and real NNN neighbors. This is the reason why we have chosen them. Upon selective disorder in one sublattice, the remaining clean sublattice will have real hopping parameters and, hence, it will be invariant under time reversal  ${\cal T}$. Both lattices support topologically non trivial phases which will be our starting point.

The Hamiltonian in both cases can be written as
\beqa
H=\sum_{{\mathbf k}\in BZ} c^+_{\bf k}\cdot h_{\bf k} c_{\bf k}, \\\nonumber
h_{{\bf k}}=2t_1[\sin k_x\sigma_x+\sin k_y\sigma_y]\\\nonumber
+[\Delta-2t_2(\cos k_x+\cos k_y)]\sigma_z.
\label{eq_Hsquare}
\eeqa
where $t_1, t_2$ are intra-inter orbital couplings respectively and $\Delta$ is a staggered potential. A schematic representation of the hopping structure of both lattices is shown in Fig. \ref{fig_square} (upper  part). 
The first example  called diatomic square lattice \cite{Hou13,ON18} is very similar to the Haldane model: The unit cell has two square plaquettes traversed by a non trivial  flux (caused in this case by the NN neighbor hoppings) of opposite  signs so that the total flux in the Brillouin zone is zero. 
The second example is that of the two orbital model on the square lattice. This  has been used as an example of a simple tight binding model supporting Chern number bigger than one \cite{PHB10,GN12}. A schematic representation of the hoppings is shown in Fig. \ref{fig_square} (upper right part). In this case the fictitious flux induced by the complex hopping parameters per plaquette is zero. Although the two lattices are  quite different, the remaining clean sublattice after selective disorder are identical. 

We have analyzed the localization behavior  of both lattices in the non--trivial topological phase under selective disorder in one sublattice only.  Interestingly, despite the non--trivial topology of the bands, these lattices undergo standard Anderson localization upon selective disorder. Their level space variances are shown in Fig. \ref{fig_square} (lower part).
There is not any region in energy where the variance reaches the GUE value, so no extended stated remain in the spectrum.   

\section{Discussion and summary}
%
\begin{center}
\begin{table}[h]
\begin{tabular}{|c || c | c | c |}
\hline
Model & \;$\Omega$ \; & ${\cal T}$ broken & Metal  \\
\hline\hline
Haldane & yes & yes & yes  \\
TI & no & yes  & yes  \\
TII & yes & yes  & yes   \\
Squared bi-orbital & yes & no & no \\
Squared diatomic  & yes &  no & no \\
\hline
\end{tabular}
\caption{Summary of the examples given in the text. $\Omega$ means that the original lattice has non--trivial 
Berry curvature. ${\cal T}$ broken refers to the clean sublattice that remains after selective disorder. }
\label{T1}
\end{table}
\end{center}
%
All the models discussed in this work -except  the TI model on the honeycomb lattice - are examples of Chern insulators defined on partite two dimensional lattices. The objective was to discuss the generality and robustness of the  band of critical states found in the Haldane model under strong selective disorder \cite{CGL16}. The results are summarized in Table I. Perhaps the  TI case,  a topologically trivial, time reversal broken model, is the most significant. Previous studies of localization (or rather {\it lack of}) established the presence and robustness of extended states carrying the Chern number in topologically non trivial models \cite{Pruisken88,Letal94,XSetal12,WBETAL15,QHetal16} in the class A (absence of any discrete symmetry). The presence of a robust metallic state in the TI lattice  leads us to conclude  that, irrespective of the topological properties of the original system (first column in table \ref{T1}),  the two minimal requirements for the absence of localization upon selective disorder are full connectivity, and time reversal invariance broken in the ``clean" sublattice.  Absence of localization in the presence of magnetic fields (broken time reversal symmetry) has been analyzed at length in the literature \cite{KB97,WSetal15,SWetal16} but the hopping structure of the TI model shown in the left hand side of Fig. \ref{fig_fluxes} does not correspond to a magnetic field. The standard localization properties of the topologically non-trivial square lattices shown in Fig. \ref{fig_square} allows us to discard topology as a key point in the absence of localization.  As mentioned in the introduction, a band of critical states has been found in models with broken ${\cal T}$ and spin rotation symmetry. The models presented in this work are spin rotation invariant (in fact they are spin--less models) but perhaps the sublattice symmetry plays a similar role as spin rotation (a ${\cal Z}_2$ symmetry in planar systems), broken by the selective disorder. 

An important remark about our results refers to the precise nature of the partite lattices considered. The simplest example  of a Chern insulator,  the Haldane model, is defined in the Honeycomb lattice which is naturally made of two interpenetrating triangular lattices. Selective disorder affects one of the sublattices. When there are internal degrees of freedom, as in the case of the two orbital model in the squared lattice, the  selective disorder considered affects an entire {\it geometrical} sublattice (it destroys both orbitals in nearest neighbour sites). 
Should we had  considered one of the orbitals, say, A (blue in Fig. 4 upper right) as a sublattice,  selective disorder of this particular sublattice would led to the  diatomic square lattice (Fig. 4 upper left), a Chern insulator with broken ${\cal T}$ and no metallic state.

The findings of this work show that the role of disorder on 2D systems, even if non--interacting,  is still far from been fully understood. The experimental advances in optical lattice realizations can shed light on the nature of the critical metal found quite systematically in our ${\cal T}$ broken examples.

\begin{acknowledgments}
We gratefully acknowledge Oscar Pozo for help with the figures. EC acknowledges the financial support of FCT-Portugal through grant No. EXPL/FIS-NAN/1728/2013. This research was supported in part by the Spanish MECD grants FIS2014-57432-P, the  European Union structural funds and the Comunidad de Madrid MAD2D-CM Program (S2013/MIT-3007).
\end{acknowledgments}
\bibliography{Metal}
\end{document}